\documentclass[a4paper,10pt]{article}
\usepackage{graphicx}
\usepackage{comment}
\usepackage{amssymb}
\usepackage{amsmath}
\usepackage{nicefrac}
\usepackage{graphicx}
\usepackage{dcolumn}
\usepackage{bm}
\usepackage{comment}
\usepackage{multirow}
\usepackage{cite}

\begin{document}

\huge

\begin{center}
Detailed Opacity Calculations for Stellar Models
\end{center}

\vspace{0.5cm}

\large

\begin{center}
Jean-Christophe Pain\footnote{jean-christophe.pain@cea.fr}, Franck Gilleron and Maxime Comet
\end{center}

\normalsize

\begin{center}
\it CEA, DAM, DIF, F-91297 Arpajon, France
\end{center}

\vspace{0.5cm}

\begin{abstract}
Radiative opacity is an important quantity in the modeling of stellar structure and evolution. In the present work we recall the role of opacity in the interpretation of pulsations of different kinds of stars. The detailed opacity code SCO-RCG for local-thermodynamic-equilibrium (LTE) plasmas is described, as well as the OPAMCDF project dedicated to the spectroscopy of LTE and non-LTE plasmas. Interpretations, with the latter codes, of several laser and Z pinch experiments in conditions relevant to astrophysical applications are also presented and our work in progress as concerns the internal solar conditions is illustrated.
\end{abstract}

\section{Introduction}
The knowledge of opacity (frequency-dependent photoabsorption cross section per unit mass) is crucial in the understanding and modeling of radiation transport. The applications encompass inertial confinement fusion (NIF/LMJ), magnetic confinement fusion (ITER, e.g., radiative losses from the divertor tungsten-coated tiles), and astrophysics. Models of stellar structure and evolution are very sensitive to radiative transfer and opacity. \cite{Frost06} discovered the variability of the radial velocity (the photosphere approaches and recedes alternatively from the observer) of $\beta$ Cephei, which has a magnitude from +3.16 to +3.27 and a period of about 4.57 hours. Spectroscopists measure the effect of the radial velocity on the absorption lines from different levels of the star's atmosphere. The opacity of the {\it iron group} (Cr, Fe, and Ni) is particularly important for the envelopes of $\beta$ Cephei ($\beta$ Canis Majoris) type stars (8 to 18 M$_{\odot}$). More precisely, acoustic modes are excited through the $\kappa$ {\it mechanism}. Such stars (for instance, $\nu$ Eridani, $\gamma$ Pegasi, $\beta$ Crucis, and $\beta$ Centauri) are hot blue-white stars of spectral class B, their temperature being $T\approx 200-300\,000$ K and density $\rho\approx 10^{-7}-10^{-6}$ g/cm$^3$ \cite{Turck2016}.

In 1924, \cite{Payne25} demonstrated the great preponderance of hydrogen and helium in stars, and in 1929, \cite{Russell29} published the first quantitative analysis of the chemical composition of the solar atmosphere. Opacity was already known as a key parameter of stellar models since \cite{Eddington26}. Although atoms are not fully ionized in such conditions, photoexcitation and photoionization were neglected at the time. In the early 1960s, \cite{Cox64} and \cite{Huebner64} introduced the two latter processes in the calculation of stellar opacities. \cite{Simon82} brought forth in 1982 the Cepheid pulsation problem (named after $\delta$ Cephei, not to be confused with $\beta$ Cepheids) and its connection to the opacity of ``heavy'' elements: C, N, O, etc. In the 1990s, two important projects were launched: OPAL (LLNL) \cite{Iglesias96} and OP (Opacity Project, an international academic collaboration) \cite{Seaton87,OP1,OP2,OP3}, which provided the second-generation stellar opacity tables.

Opacity drives the pulsations of many variable stars. In cases where the opacity increases with temperature, the atmosphere becomes unstable with pulsations that are governed by the $\kappa$ mechanism, which consists of the following steps.
\begin{itemize}
  \item The inward motion of a layer tends to compress the layer and increase the density
  \item The layer becomes more opaque and the flux from the deeper layers gets stuck in the high-opacity region
  \item The heat increase causes a pressure buildup pushing the layer back out again
  \item The layer expands, cools, and becomes more transparent to radiation
  \item Energy and pressure beneath the layer diminish
  \item The layer falls inward and the cycle repeats itself.
\end{itemize}

It is worth mentioning that {\it slowly pulsating B} (SPB) stars are subject to gravity modes (g modes) also connected to the iron opacity. Their mass varies from 2 to 6~M$_{\odot}$.

The detailed opacity code SCO-RCG for local-thermodynamic-equilibrium (LTE) plasmas is described in Section~\ref{sec2}, as well as the OPAMCDF project \cite{Comet17a,Comet17b} dedicated to the spectroscopy of LTE and non-LTE plasmas. The interpretations of some laser and Z-pinch experiments are presented in Section~\ref{sec3}, and the modeling of the Sun interior is discussed in Section~\ref{sec4}.

\section{Computation of Atomic Data and Radiative Opacity}\label{sec2}

\subsection{The Different Processes in LTE Plasmas}

Opacity is the sum of photoexcitation (bb: bound--bound), photoionization (bf: bound--free), and inverse Bremstrahlung (ff: free--free) corrected by the stimulated-emission effect $\left(1-\exp(-h\nu/k_BT)\right)$, and scattering (s) contributions:
\begin{equation}
\kappa(h\nu)=\left[\kappa_{\mathrm{bb}}(h\nu)+\kappa_{\mathrm{bf}}(h\nu)+\kappa_{\mathrm{ff}}(h\nu)\right]
\left[1-\exp(-h\nu/k_BT)\right]+\kappa_s(h\nu)\ .
\end{equation}
Photoexcitation can be described as
\begin{equation}
X_i^{q+}+h\nu \rightarrow X_j^{q+}\ ,
\end{equation}
where $X_i^{q+}$ is an ion with charge $q$ in an excitation state $i$. The signature of the absorbed photon $h\nu$ is a spectral line. The corresponding opacity contribution reads
\begin{equation}
\kappa_{\mathrm{bb}}(h\nu)=\frac{1}{4\pi\epsilon_0}\frac{\mathcal{N}_A}{A}\frac{\pi e^2h}{m_ec}\sum_{i\rightarrow j}\mathcal{P}_if_{ij}\Psi_{ij}(h\nu)\ ,
\end{equation}
where $\mathcal{P}_i$ is the population of initial level $i$, $f_{ij}$ the oscillator strength, and $\Psi_{ij}$ the profile of the spectral line corresponding to the transition $i\rightarrow j$ accounting for broadening mechanisms (Doppler, Stark, etc.). Also $\epsilon_0$ is the dielectric constant, $\mathcal{N}_A$ Avogadro's number, $e$ and $m_e$ represent respectively the electron charge and mass, $c$ is the speed of light, and $A$ the atomic mass of the considered element. Photoionization is a threshold process that occurs when a bound electron $e^-$ is ejected after absorption of a photon with high enough energy,
\begin{equation}
X_i^{q+}+h\nu \rightarrow X_j^{(q+1)+}+e^-\ .
\end{equation}

Bremsstrahlung refers to the radiation emitted by an electron slowing down in the electromagnetic field of an ion. The inverse process occurs when a free electron and an ion absorb a photon
\begin{equation}
h\nu+\left[X_i^{q+}+e^-(\epsilon)\right]\rightarrow X_j^{q+}+e^-\left(\epsilon'\right)\ ,
\end{equation}
$\epsilon$ and $\epsilon'$ being the energies of the free electron before and after photoabsorption. Calculations of the free--free cross section involve quantities related to the elastic-scattering matrix elements of ionic electron-impact excitation. The scattering of a photon by a free electron can be accounted for using the Klein--Nishina  cross section (we set $\gamma=h\nu/m_ec^2$)
\begin{equation}
\sigma_{\mathrm{KN}}(h\nu)=2\pi r_0^ 2\left(\frac{1+\gamma}{\gamma^3}\left[\frac{2\gamma(1+\gamma)}{1+2\gamma}-\ln(1+2\gamma)\right]+
\frac{\ln(1+2\gamma)}{2\gamma}-\frac{(1+3\gamma}{(1+2\gamma)^2}\right)\ ,
\end{equation}
where $r_0=e^2/(m_ec^2)$ is the classical radius. If $\gamma\ll 1$, $\sigma_{\mathrm{KN}}(h\nu)\approx 8\pi r_0^2/3$ which is the Thomson cross section. In the so-called diffusion approximation, the radiative transfer is very sensitive to the Rosseland mean opacity, which is the harmonic mean opacity averaged over the derivative of the Planck function with respect to the temperature:
\begin{equation}\label{eqr}
\frac{1}{\kappa_R}=\int_0^{\infty}\frac{W_R(u)}{\kappa(u)}du\;\;\;\;\mathrm{where}\;\;\;\; u=\frac{h\nu}{k_BT}\;\;\;\; \mathrm{and}\;\;\;\; W_R(u)=\frac{15u^4e^{-u}}{4\pi^4\left[1-e^{-u}\right]^2}\ .
\end{equation}

\subsection{SCO-RCG Code}

The detailed (fine-structure) opacity code SCO-RCG \cite{Pain15} enables the computation of precise opacities for the calculation of accurate Rosseland means (see Eq. (\ref{eqr})). The (super-)configurations are generated on the basis of a statistical fluctuation theory and a self-consistent computation of atomic structure is performed for all the configurations. Thus each configuration has its own set of wave functions determined in a single-configuration approximation. A peculiarity of the code is that it does not rely on the ``isolated atom'' picture but on a realistic atom-in-plasma model (such as used for equation-of-state calculations). Relativistic effects are taken into account with the Pauli approximation. The {\it detailed line accounting} (DLA) part of the spectrum is performed using an adapted version of the RCG routine from the suite of atomic structure and spectra codes of \cite{Cowan81}. The RCG source code has been used by spectroscopists for decades, it has many available options, and is well documented. In SCO-RCG criteria are defined to select the transition arrays that can be treated line-by-line. The data required for the calculation of the detailed transition arrays (Slater, spin--orbit, and dipolar integrals) are obtained from SCO, thus providing a consistent description of the plasma screening effects on the wave functions. Then the level energies and the lines are calculated by RCG.

The computation starts with an average-atom calculation in LTE, which provides the average populations of the subshells. 
A superconfiguration is made of supershells (a supershell being a group of subshells) populated by electrons. In SCO-RCG, all the supershells are ordinary subshells, except one (the so-called Rydberg supershell), which gathers all the highly-excited subshells. We use LTE fluctuation theory around the average-atom, non-integer populations in order to determine the range of the population variations and, therefore, the possible list of configurations (if the Rydberg supershell is empty) or superconfigurations (if the Rydberg supershell is occupied). The superconfigurations are then sorted out according to their Boltzmann weights, estimated using the average-atom wave functions, only keeping the (super-)configurations having the highest weights.

The strength of this approach is that it enables us to take into account many highly-excited states and satellite lines. The populations of those states may be small but their number is so huge that they can play a significant role in the opacity. The orbitals in the Rydberg supershell are chosen so they weakly interact with the inner orbitals. A DLA calculation is performed if possible and necessary for all the transition arrays starting from the configuration; DLA computations are carried out only for pairs of configurations giving rise to less than 800\,000 lines. In other cases, transition arrays are represented by Gaussian profiles in the {\it unresolved transition array} (UTA) \cite{Bauche-Arnoult79,Bauche-Arnoult82} or {\it spin--orbit split array} (SOSA) \cite{Bauche-Arnoult85} formalisms. If the Rydberg supershell contains at least one electron, then transitions starting from the superconfiguration are taken into account by the {\it super transition array} (STA) model \cite{Bar89}. The amount of detailed calculations performed in SCO-RCG is now largely dominant.

The {\it partially resolved transition array} (PRTA) model \cite{Iglesias12} has been recently implemented. It enables us to replace many statistical transition arrays by small-scale DLA calculations obtained by removing the passive subshells from the ``real'' configuration to form the reduced configuration. The DLA computation of the reduced configuration is performed using the wave functions of the ``real'' configuration previously calculated. The electrostatic variance due to the passive subshells is added to each line of the DLA calculation in order to keep constant the total oscillator strength of the transition array. We have extended this approach to the STA formalism of \cite{Bar89}, omitting the Rydberg supershell in the computation, and adding its contribution to the widths of all lines. The contribution of the Rydberg supershell is included as a Gaussian ``dressing function'' \cite{Pain15}. We also have the possibility of replacing this dressing function by a coarse-grain configurationally resolved profile, following the {\it configurationally resolved super transition array} (CRSTA) method of \cite{Kurzweil16}.

\subsection{The OPAMCDF Project}

We have recently started a project, referred to as OPAMCDF, dedicated to the spectroscopy of LTE and NLTE plasmas using the multi-configuration Dirac--Fock (MCDF) code \cite{Bruneau83,Bruneau84}. The code involves methods similar to the ones published by Grant \cite{Grant70,Grant07}.

\subsubsection{Opacity Calculation}

To generate the list of configurations that have to be calculated \cite{Comet15}, a {\it relativistic average atom model} (RAAM) is used in order to determine average plasma quantities such as the average ionization and average occupation of each subshell. The population variance of each subshell is estimated using a binomial function assuming uncorrelated electrons. For the calculation of the photoexcitation opacity, subshells are divided into two parts: the first contains subshells with principal quantum number in the range $1\leq n\leq 6$ and the second subshells with principal quantum number $n\geq 7$. The transition arrays between two configurations with no electron in the second part are treated either in DLA (if the number of lines in the transition array is $\leq 2\times 10^6$) or PRTA \cite{Iglesias12} if it is larger than $2\times 10^6$. In the latter case, subshells are sorted out according to their contribution to the total variance of the transition array. The criterion is based on the number of lines: subshells with the smallest contribution to the variance of the transition array are removed until the number of lines of the reduced DLA calculation is lower than $2\times 10^6$. \cite{Iglesias12} remove the passive subshells from the ``real'' configuration whichever their population. In fact, this can be improved by taking into account that the electrons in a subshell can be removed ``one by one'' \cite{Comet17b}. On the other hand, if there is at least one electron in the second part (highly-excited configurations), transition arrays are calculated in the SOSA formalism, relying on an efficient direct computation of the two-electron relativistic strength-weighted mean and variance of the line energies \cite{Krief15}.

For each transition array of the lines between two configurations, a MCDF calculation is performed in the framework of the so-called Slater's {\it transition state} approximation \cite{Godefroid76, Bruneau83}. It ensures good accuracy in terms of the line energies and orthogonality of the initial and final wave functions. All MCDF calculations include the Breit interaction, QED corrections (self-energy, vacuum polarization), and nucleus effects (finite nuclear mass and recoil). Moreover, wave functions take into account the finite size of the nucleus using a Fermi (or Woods--Saxon) distribution. In the present version of the code, the photoionization opacity is calculated in the {\it detailed configuration accounting} approximation.

\subsubsection{Cross Sections and Rates for NLTE Modeling}

For the spectroscopy of non-LTE plasmas, cross sections and rates are needed for all atomic processes that can populate or depopulate atomic states, levels, configurations, or superconfigurations. In addition to photoexcitation, the MCDF code can be used to calculate cross sections and rates for photoionization, autoionization, collisional excitation, and ionization using the {\it distorted wave} approximation (DW) or the close-coupling method (except for collisional ionization). Originally the code was able to calculate rates and cross sections only between levels. Recent modifications have been carried out in order to implement the configuration-averaged approximation for all processes listed above. For collisional processes, the Breit (B), generalized Breit (GB), and, more recently, the generalized Breit plus the imaginary part of the matrix element (GBI) can be included in the collisional matrix elements \cite{Sampson09}. For autoionization, the Breit interaction can also be included. Table~\ref{Xe_CS} lists collisional strengths for Xe$^{53+}$ at a scattered-electron energy of 112~keV. The experimental values were obtained with an electron beam ion trap (EBIT) \cite{Widmann00,Chen08}. In Table~\ref{Fe+Mo_ICS}, collisional ionization cross sections for Fe$^{25+}$ and Mo$^{41+}$ are tabulated for different electron impact energies. All values are compared to measurements performed on the EBIT by \cite{Watanabe03}. These comparisons show that our results are very close to the experimental values.

\begin{table}[!ht]
  \caption{Comparison of collisional excitation cross sections (barns) for H-like Xe$^{53+}$ at an electron energy of 112~keV.}\label{Xe_CS}
  \smallskip
  {\centering\small
  \begin{tabular}{lcccc}  
  \noalign{\smallskip}\hline
  Line & Experiment$^{\rm a}$ & MCDF(C)$^{\rm b}$ & MCDF(C+B)$^{\rm c}$ & MCDF(C+GBI)$^{\rm d}$ \\\hline
  \noalign{\smallskip}
  \noalign{\smallskip}
  Ly$_{\alpha 1}$ & 8.6 $\pm$ 1.5 & 8.253 & 7.318 & 8.042\\
  Ly$_{\alpha 2,3}$ & 8.2 $\pm$ 3.4 & 6.567 & 6.401 & 6.717\\\hline
  \noalign{\smallskip}
  \end{tabular}

  }

  \smallskip
  \scriptsize
  \noindent
  \ \ \ \ \ \ \ \ \ \ \ \ \ \ \ \ \ \ {\bf Notes.}
  
  \ \ \ \ \ \ \ \ \ \ \ \ \ \ \ \ \ \  $^{\rm a}$Electron beam ion trap \cite{Widmann00,Chen08}.
   
  \ \ \ \ \ \ \ \ \ \ \ \ \ \ \ \ \ \  $^{\rm b}$MCDF with the Coulomb interaction. 
  
  \ \ \ \ \ \ \ \ \ \ \ \ \ \ \ \ \ \  $^{\rm c}$MCDF with the Coulomb and Breit interactions. 
  
  \ \ \ \ \ \ \ \ \ \ \ \ \ \ \ \ \ \  $^{\rm d}$MCDF with the Coulomb and GBI interactions. 
  
  \ \ \ \ \ \ \ \ \ \ \ \ \ \ \ \ \ \  The Ly$_{\alpha 1}$ line corresponds to ${\rm 1s}\ ^2{\rm S}_{1/2}\rightarrow {\rm 2p}\ ^2{\rm P}_{3/2}$ and Ly$_{\alpha 2,3}$ to the lines ${\rm 1s}\ ^2{\rm S}_{1/2}\rightarrow {\rm 2p}\ ^2{\rm P}_{1/2}$ and ${\rm 1s}\ ^2{\rm S}_{1/2}\rightarrow {\rm 2s}\ ^2{\rm S}_{1/2}$.
\end{table}

\begin{table}[!ht]
  \caption{Comparison of collisional ionization cross sections (in units of 10$^{-22}$~cm$^2$) for H-like Fe$^{25+}$ and Mo$^{41+}$.}\label{Fe+Mo_ICS}
  \smallskip
  {\centering\small
  \begin{tabular}{lccc}  
  \noalign{\smallskip}\hline
  & $E_e$ (keV) & Experiment$^{\rm a}$  & MCDF(C)$^{\rm b}$\\\hline
  \noalign{\smallskip}
  Fe$^{25+}$ & 13.3 & 0.93 $\pm$ 0.24 & 1.33\\
             & 14.8 & 1.30 $\pm$ 0.12 & 1.56\\
             & 17.3 & 1.51 $\pm$ 0.15 & 1.80\\
             & 19.8 & 1.69 $\pm$ 0.23 & 1.95\\
             & 24.8 & 1.75 $\pm$ 0.34 & 2.10\\
             & 29.8 & 2.08 $\pm$ 0.79 & 2.15\\
             & 39.6 & 2.12 $\pm$ 1.18 & 2.17\\\hline
  Mo$^{41+}$ & 49.4 & 0.282 $\pm$ 0.022 & 0.281\\
             & 64.4 & 0.313 $\pm$ 0.029 & 0.314\\
             & 79.6 & 0.323 $\pm$ 0.051 & 0.329\\\hline
 \noalign{\smallskip}
  \end{tabular}

  }
  \smallskip
  \scriptsize
  \noindent
  \ \ \ \ \ \ \ \ \ \ \ \ \ \ \ \ \ \ \ \ \ \ \ \ \ \ \ \ \ \ \ \ \ \ \ \ \ \ \ \ \ \ {\bf Notes.} $^{\rm a}$EBIT \cite{Watanabe03}. $^{\rm b}$MCDF with the C.
\end{table}

The implementation of configuration-averaged cross sections and rates gives us the opportunity to check the validity of semiempirical formulae that are widely used for collisional--radiative modeling. The main differences between SCO-RCG and OPAMCDF are summarized in Table~\ref{dif}.

\begin{table}[!ht]
  \caption{Main differences between SCO-RCG and OPAMCDF.}\label{dif}
  \smallskip
  {\centering\small
  \begin{tabular}{lll}  
  \noalign{\smallskip}\hline
  & SCO-RCG & OPAMCDF \\
  \noalign{\smallskip}
  \noalign{\smallskip}\hline
  Hamiltonian & Schr\"odinger + Pauli & Dirac\\\hline
  Density effects on WF & Yes & No\\\hline
  Exchange & Exact & \cite{Ichimaru87}\\\hline
  Relaxation & Yes & Transition state\\\hline
  Breit + QED & No & Yes\\\hline
  No. of lines per array & $8\times 10^5$ & $2\times 10^6$\\\hline
  Calculation time & 1 day & A few days\\\hline
  Statistical methods & UTA, SOSA, PRTA, STA, CRSTA & SOSA, PRTA\\\hline
  Rosseland mean & Yes & Not yet\\\hline
  \noalign{\smallskip}
  \end{tabular}

  }
\end{table}

\section{Interpretation of Spectroscopic Experiments}\label{sec3}

\subsection{Laser Experiment}

\begin{figure}
\begin{center}
\includegraphics[width=.6\textwidth]{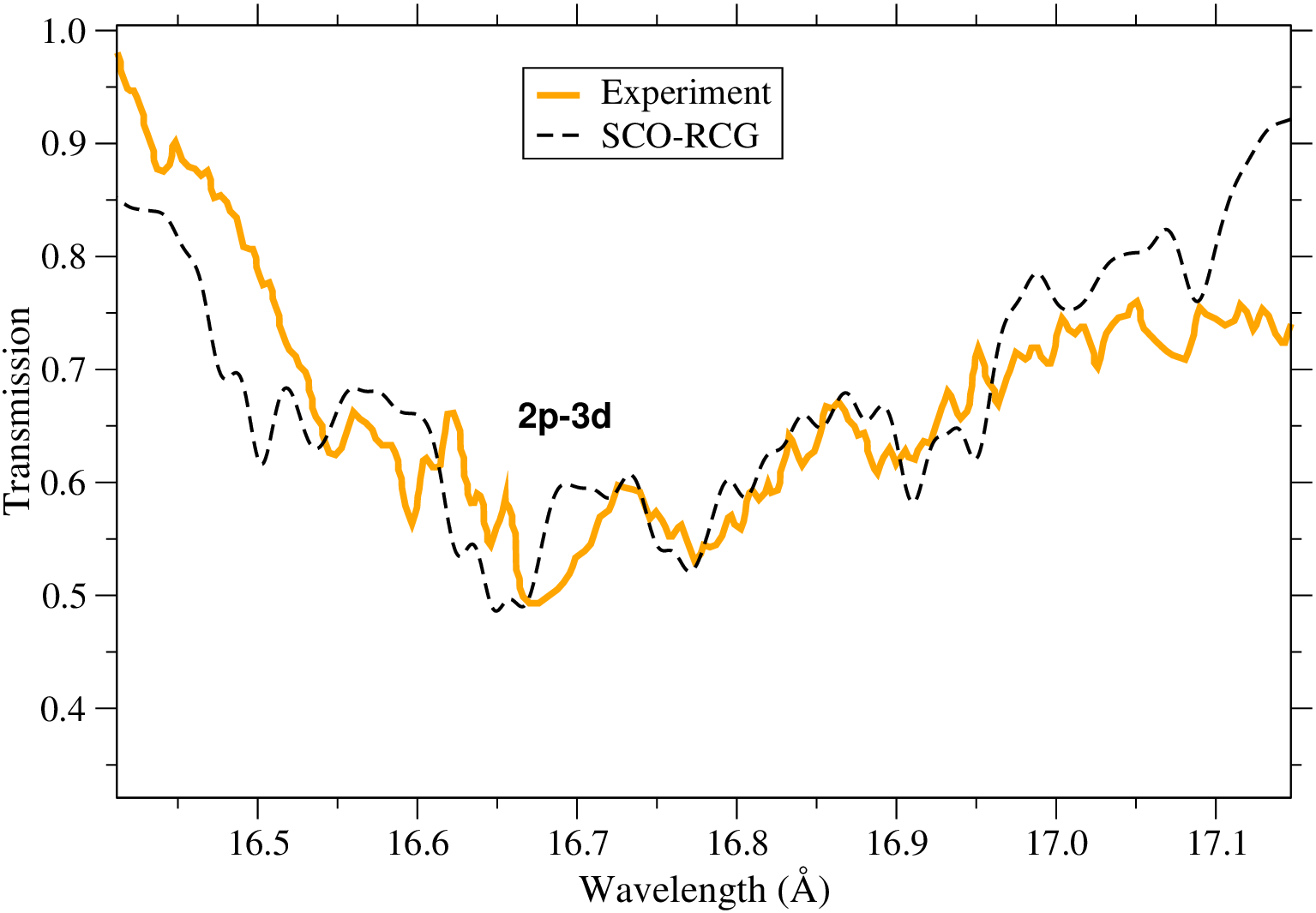}\label{chenais_2}
\caption{Comparison between an experimental transmission spectrum of Fe measured on the ASTERIX IV laser facility \cite{Chenais00} and a SCO-RCG calculation at $T=20$~eV and $\rho=0.004$~g/cm$^{3}$. The areal mass is 8 $\mu$g/cm$^2$ and the resolving power $E/\Delta E=300$.}
\end{center}
\end{figure}

Fig.~\ref{chenais_2} displays an interpretation with SCO-RCG at $T=20$~eV and $\rho=0.004$~g/cm$^{3}$ of a transmission spectrum of Fe measured on the ASTERIX IV laser facility by \cite{Chenais00}. The spectrum corresponds to a part of a spectrum of the quasar IRAS 13349+2438 measured by the XMM-Newton observatory for the ions Fe vii--Fe xii. Although the main absorption structures are reproduced, the agreement is not perfect regarding the transition energies and transmission levels. The discrepancies might be attributed to temperature and density gradients and to uncertainties in the knowledge of the areal mass.

\vspace{3mm}

\vspace{4mm}

\begin{figure}
\begin{center}
\includegraphics[width=.6\textwidth]{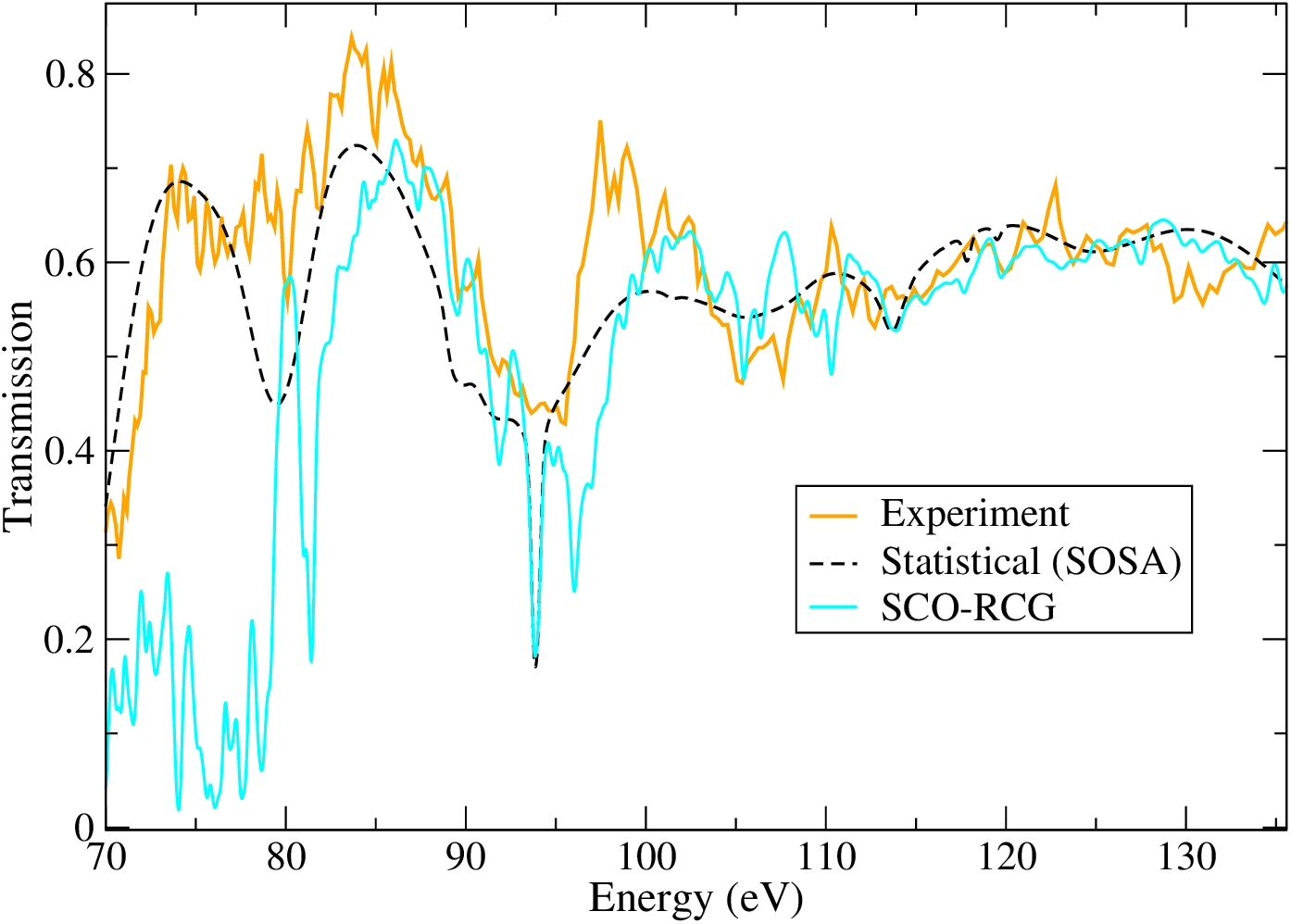}\label{eidmann_1}
\caption{Comparison between an experimental transmission spectrum of Fe measured on the ASTERIX IV laser facility \cite{Winhart96} and calculations at $T=22$~eV and $\rho=0.01$~g/cm$^{3}$. The areal mass is 15 $\mu$g/cm$^2$ and the resolving power $E/\Delta E=300$.}
\end{center}
\end{figure}

Fig.~\ref{eidmann_1} presents an interpretation at $T=22$~eV and $\rho=0.01$~g/cm$^{3}$ of a transmission spectrum of Fe measured on the ASTERIX IV laser facility by \cite{Winhart96} in the XUV range where configuration interaction is important. In SCO-RCG only configuration interaction between relativistic subconfigurations of a non-relativistic configuration is included. It may be seen that we obtain better agreement with a statistical calculation (SOSA) than with a DLA computation (SCO-RCG). This point, which might be fortuitous, is unexplained; however, \cite{Turck2015} observed such a behavior in other measurements and as yet it remains a mystery.

\begin{figure}
\begin{center}
\includegraphics[width=.6\textwidth]{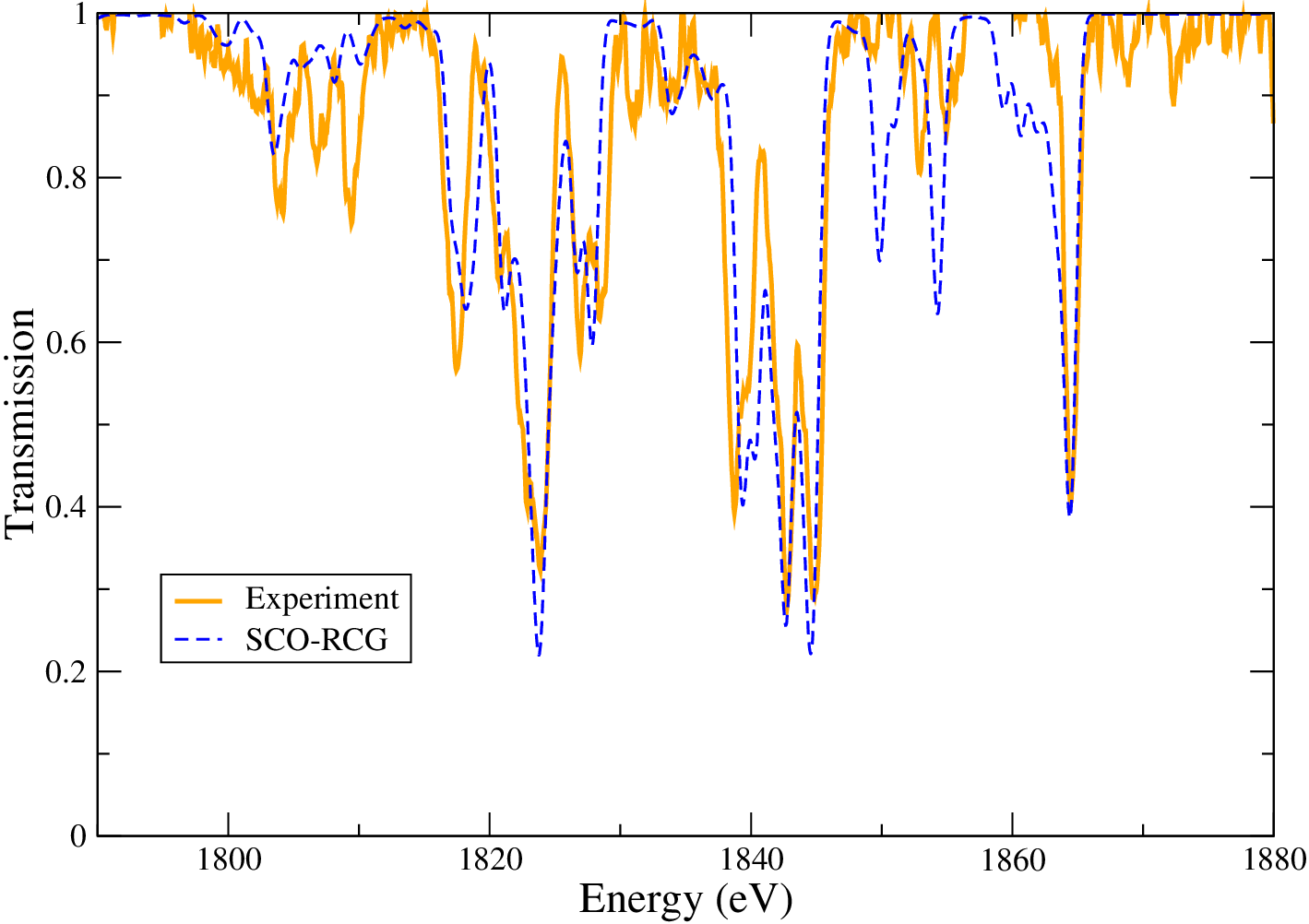}\label{Si}
\caption{Comparison of an experimental transmission spectrum of Si measured on the SG II laser facility \cite{Xiong16} and a spectrum computed by SCO-RCG at $T=72$~eV and $\rho=0.006$~g/cm$^{3}$. The areal mass is 23~$\mu$g/cm$^2$ and the resolving power $E/\Delta E=2000$.}
\end{center}
\end{figure}

In the last decades, K-shell absorption lines of silicon ions have been extensively observed in various astrophysical objects with the high-resolution spectrometers of the XMM-Newton, Chandra, and Suzaku space missions. Silicon absorption lines were observed in active galactic nuclei. Silicates are an important component of cosmic matter that are formed in the winds of AGB (asymptotic giant branch) stars (evolved, cool, luminous stars). Silicates are processed in the diffuse interstellar medium and are also present in dust of protoplanetary disks. The comparison between a recent experiment by \cite{Xiong16} (see Fig.~\ref{Si}) for the ions Si ix--Si xiii and SCO-RCG calculation shows good agreement except around $h\nu=1855$~eV. The differences are expected to be due to configuration interaction (between non-relativistic configurations) excluded in the computation (investigation is currently in progress).

\vspace{5mm}

\subsection{Z-pinch Experiments}

\cite{Bailey07} reported iron transmission measurements at $T= 156$~eV and $N_e=6.9\times 10^{21}$~cm$^{-3}$ over the photon energy range $h\nu\approx 800{-}1800$~eV. The samples consisted of an Fe/Mg mixture fabricated by depositing 10 alternating Mg and Fe layers, fully tamped on both sides by a 10~$\mu m$ thick parylene-N (C$_8$H$_8$). The challenges of high-temperature opacity experiments in this work were overcome by using the dynamic hohlraum X-ray source of the Z facility at the Sandia National Laboratory (SNL). The process entails accelerating an annular tungsten Z-pinch plasma radially inward onto a cylindrical low-density CH$_2$ foam, launching a radiating shock propagating toward the cylinder axis. Radiation trapped by the tungsten plasma forms a hohlraum, and a sample attached on the top diagnostic aperture is heated during $\approx 9$~ns when the shock is propagating inward and the radiation temperature rises above 200~eV. Radiation at stagnation is used to probe the sample. The experimental spectrum was well reproduced by many fine-structure opacity codes (see Fig.~4 for a comparison with SCO-RCG and OPAMCDF), but the features around 12.4~\AA\ were not reproduced by any of them.

\clearpage

\begin{figure}
\begin{center}
\includegraphics[width=.7\textwidth]{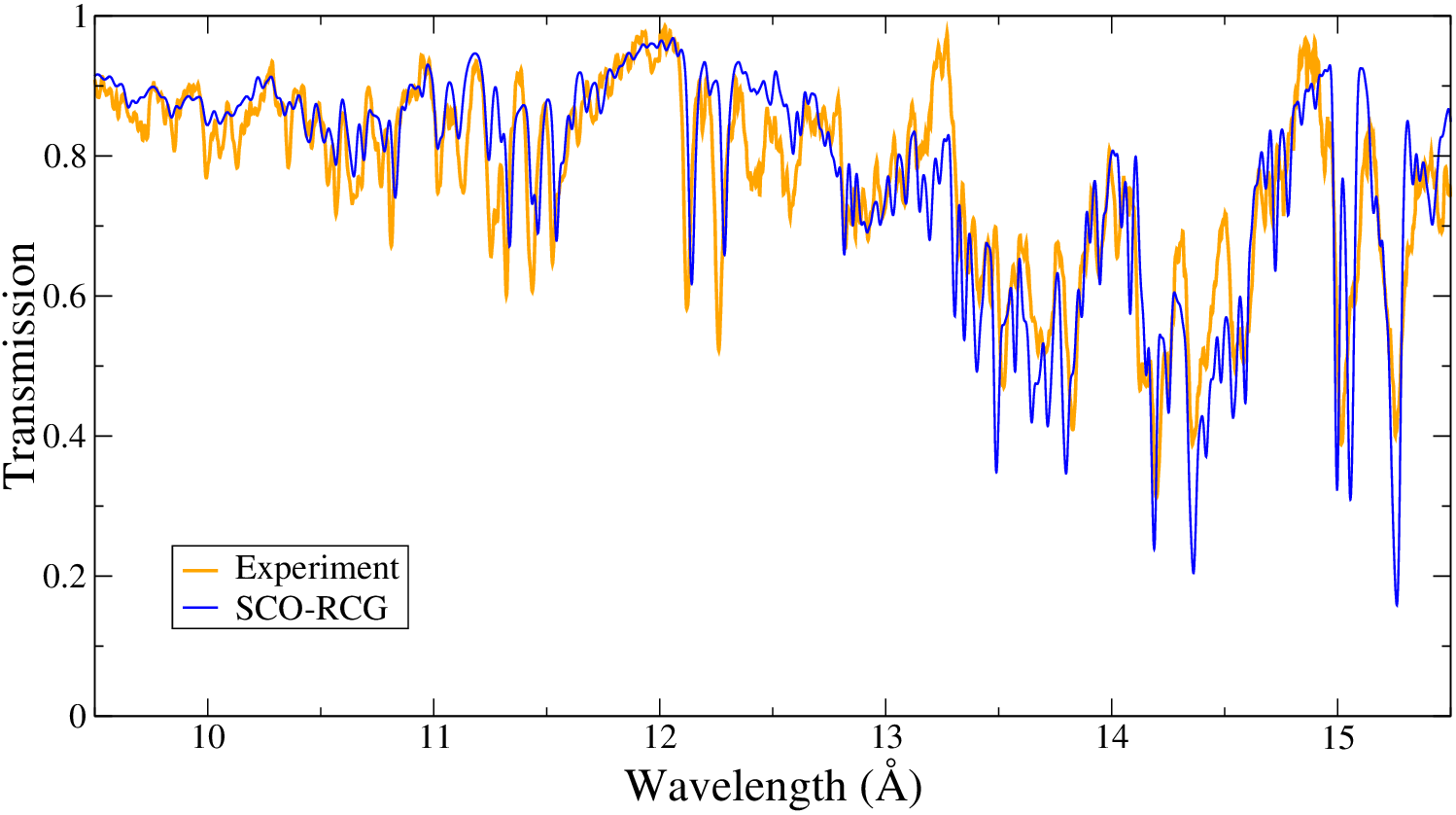}\label{figure4_new}
\caption{Comparison between the experimental spectrum (in orange) of \cite{Bailey07} and the SCO-RCG computation at $T= 150$~eV and $\rho=0.058$~g/cm$^{-3}$ (in blue).}
\end{center}
\end{figure}

\begin{figure}
\begin{center}
\includegraphics[width=.7\textwidth]{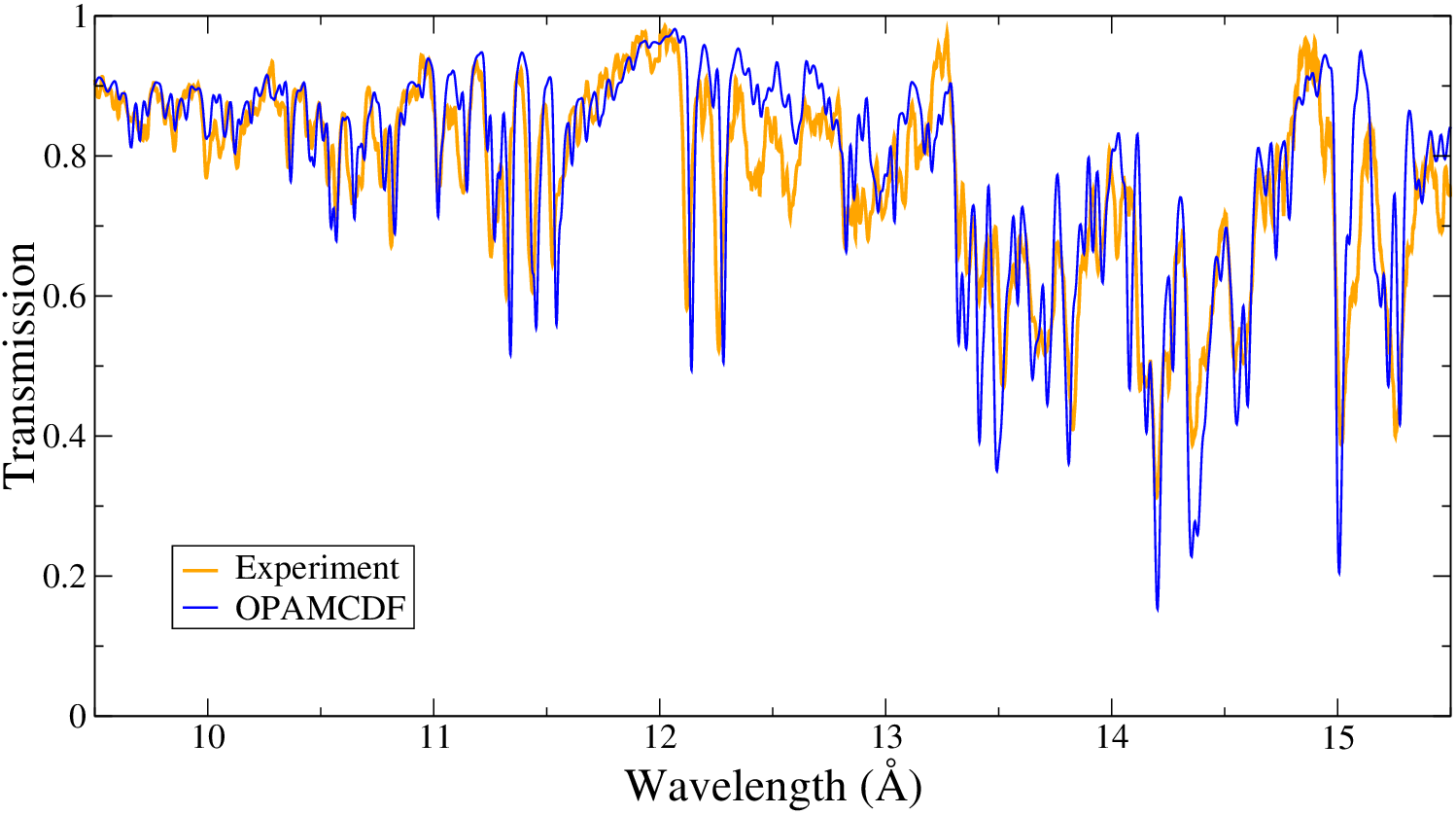}\label{figure5_new}
\caption{Comparison between the experimental spectrum (in orange) of \cite{Bailey07} and the OPAMCDF computation at $T= 156$~eV and $\rho=0.058$~g/cm$^{-3}$ (in blue).}
\end{center}
\end{figure}

\clearpage

\section{Opacity in the Interior of the Sun: Preparation of New Tables}\label{sec4}

Solar-type radiative interiors correspond to $T>2\times 10^6$~K and $\rho\approx 0.3{-}150$~g/cm$^3$. Iron contributes 25\% of the total opacity at the boundary of the convection zone (BCZ) of the Sun. The recent reevaluation of the abundances of C, N, and O in the solar mixture \cite{Asplund09} exacerbated the disagreement between helioseismic measurements and the predictions of the Standard Solar Model (SSM) \cite{Turck11}. In order to reconcile observations with model output, a 5--20\% of the opacity would be necessary. The Fe opacity recently measured on the Z machine by \cite{Bailey15} at $T=182$~eV and $N_e=3.1\times 10^{22}$~cm$^{-3}$ (conditions close to the BCZ) is a factor of 2 higher than all the theoretical spectra \cite{Pain17}. This unexplained experimental outcome is certainly stimulating new developments in the codes (photoionization, highly-excited states, two-photon processes, etc.).

\vspace{4mm}

\begin{figure}
\begin{center}
\includegraphics[width=.6\textwidth]{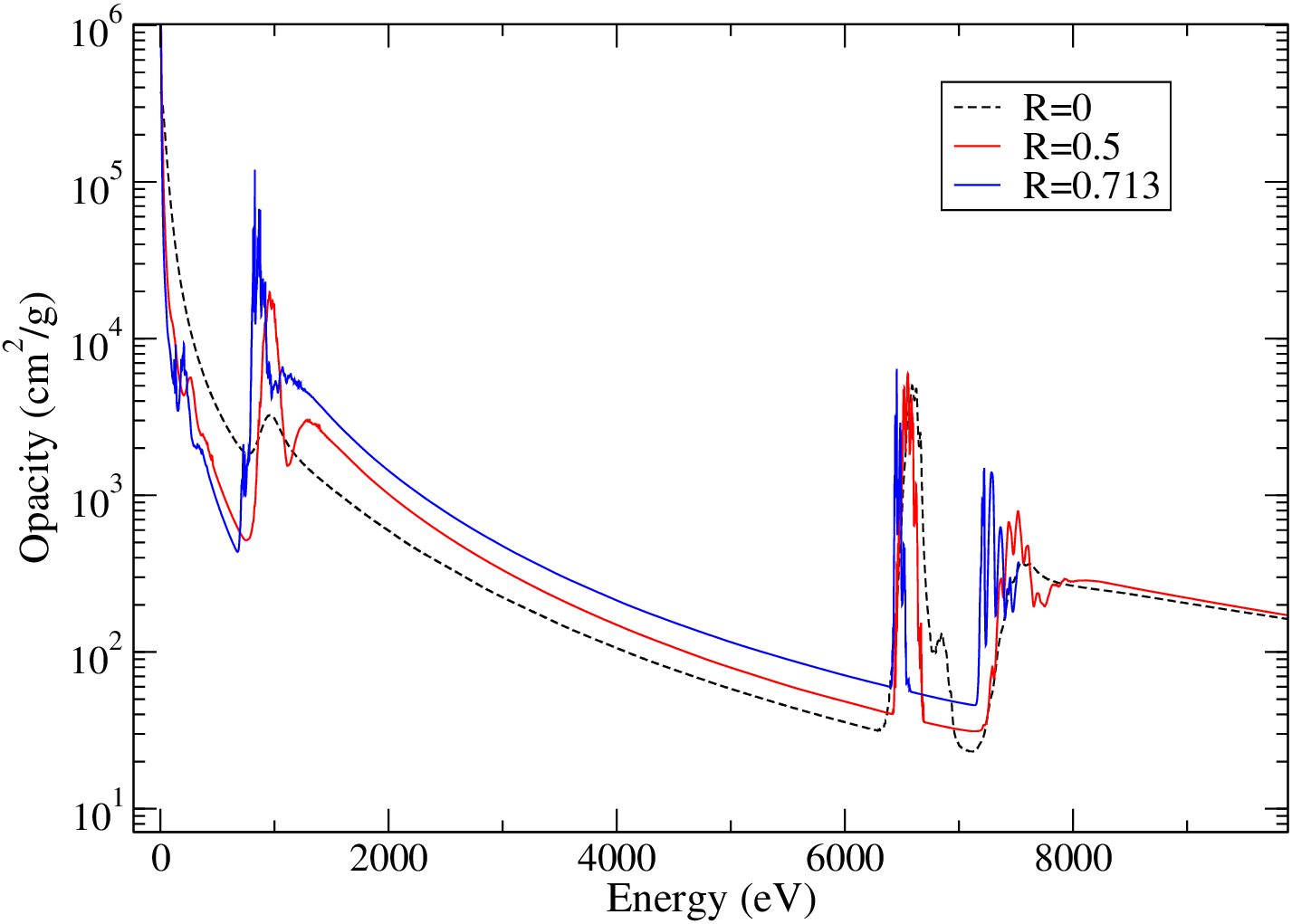}\label{Sun1}
\caption{Iron opacity at different distances (computed with SCO-RCG code) from the solar center ($R$ is expressed in unit of the solar radius $R_{\odot}$).}
\end{center}
\end{figure}

We are currently computing opacity tables in order to provide new opacities for astrophysicists. The advantages of SCO-RCG when compared to other codes that have been used up to now to produce stellar opacity tables are: a proper modeling of density effects; a precise treatment of the Stark effect for few-electron atoms; and the accounting for a huge number of excited states. Fig. 5 shows the iron opacity at different locations in the solar interior, from the core to the boundary of the convective zone. Fig.~6 displays the opacity of elements of the solar mixture of \cite{Asplund09} computed with SCO-RCG code at $T=192.92$~eV and $N_e=10^{23}$~cm$^{-3}$ in conditions similar to the BCZ. Since we are only preparing the code, in order to ensure robustness to produce tables, we did not use our new accurate modeling of Stark-effect \cite{Pain16} for those preliminary calculations. However, since a proper description of the K-shell lineshapes (e. g. for oxygen) is important, in particular for the line wings, we plan to include it.

\vspace{5mm}

\begin{figure}
\begin{center}
\includegraphics[width=.6\textwidth]{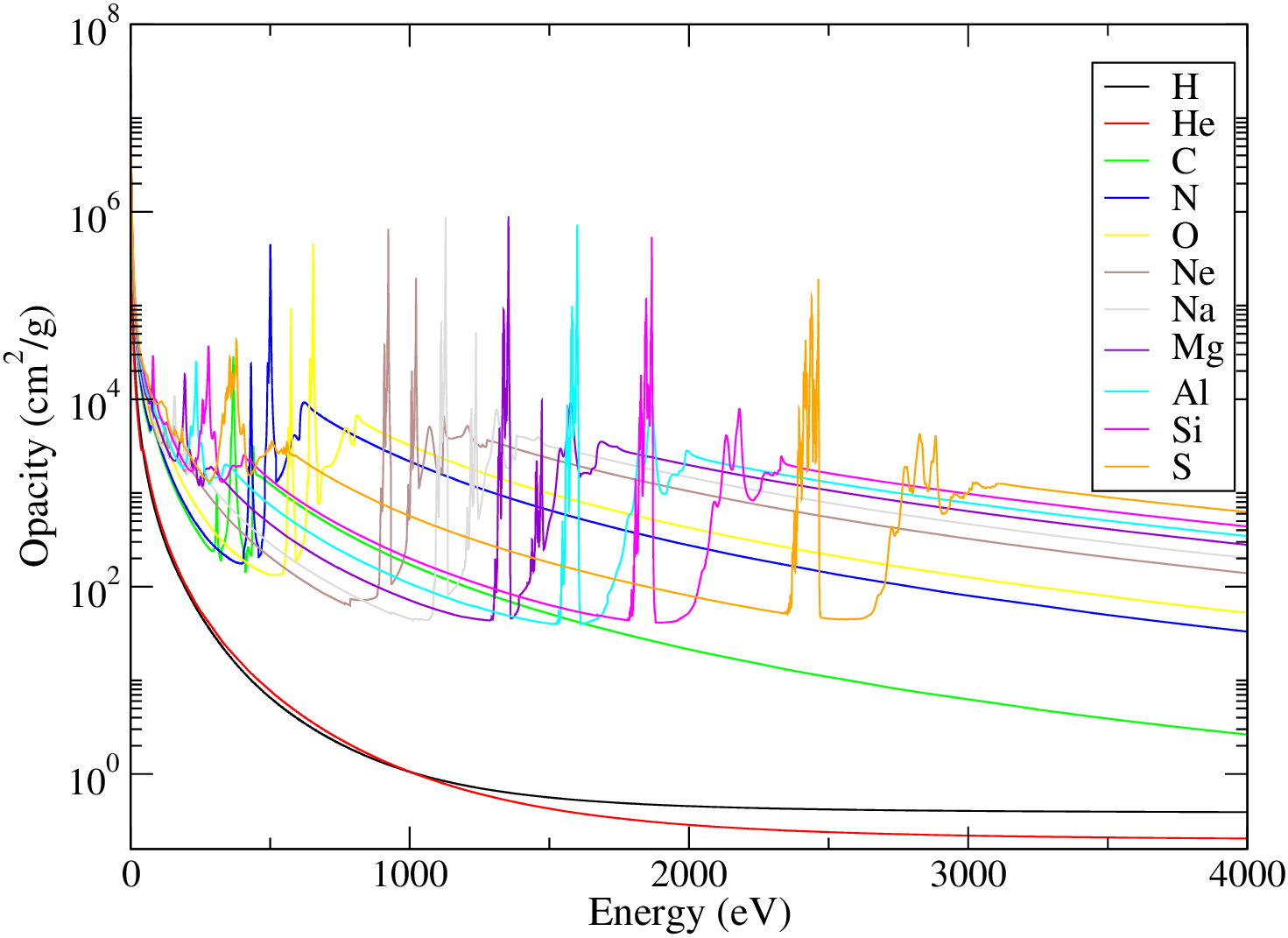}\label{figure7_new}
\caption{Opacity of elements of the solar mixture of \cite{Asplund09} at $T=192.92$~eV and $N_e=10^{23}$~cm$^{-3}$ (conditions close to the BCZ). Spectral opacities of the first 11 elements (according to atomic number $Z$).}
\end{center}
\end{figure}

\begin{figure}
\begin{center}
\includegraphics[width=.6\textwidth]{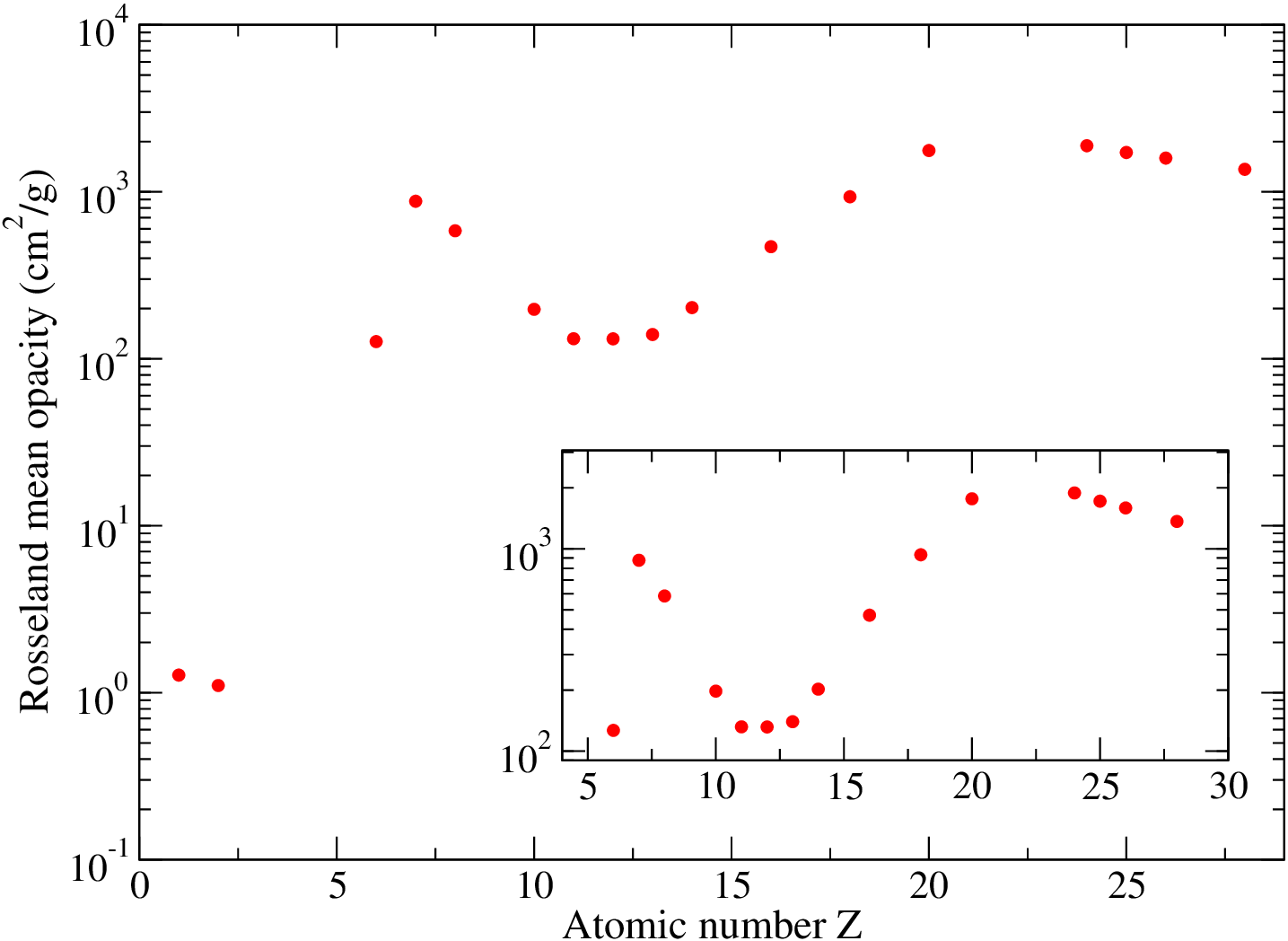}\label{figure8_new}
\caption{Opacity of elements of the solar mixture of \cite{Asplund09}. Rosseland means of the 17 elements of the mixture.}
\end{center}
\end{figure}

\clearpage

\section{Conclusion, Work in Progress and Perspectives}

Stellar models are very sensitive to the radiative opacity, and therefore, opacity computations require to take into account a huge number of levels and spectral lines. Effects such as configuration interaction are still difficult to take into account properly. An increase of 5--20\% of the opacity in the solar interior would reconcile predictions and observations. Laser or Z-pinch experiments are being performed to test the models. We are working on the solar mixture (computation of tables) as well as on associated topics such as non-LTE effects or the modeling of the Stark effect. We are also currently developing a new line-shape code (named ZEST: ZEeman-STark) \cite{Gilleron18} which will be useful for many astrophysical applications; for instance, to interpret the measurements of H$_{\beta}$, H$_{\gamma}$, and H$_{\delta}$ lines in white-dwarf photospheric/atmospheric conditions \cite{Falcon13}.

\section{Acknowledgments}

The authors would like to thank the experimental groups for providing their spectra. J.-C. Pain would like to thank the organizers, Claudio Mendoza and Manuel Bautista, for producing an event that gathered astrophysicists and specialists of the calculation of hot-plasma opacities.

\end{document}